\documentclass{article}
\usepackage[a4paper, total={5in, 8in}]{geometry}
\usepackage[utf8]{inputenc}
\usepackage{quattrocento}
\usepackage{textcomp,marvosym}
\usepackage[T1]{fontenc}
\usepackage[british]{babel}
\usepackage{amsmath}
\usepackage{amsfonts}
\usepackage{amssymb}
\usepackage{graphicx}
\usepackage{caption}
\usepackage{enumerate}
\usepackage{indentfirst}
\usepackage{xcolor}
\usepackage{url}
\usepackage{microtype}
\usepackage{tabularx}
\usepackage{appendix}
\usepackage{booktabs}
\usepackage{multirow}
\usepackage{float}
\usepackage{lipsum}
\usepackage{soul}
\usepackage{fancyhdr}
\usepackage{enumitem}
\usepackage{csquotes}

\newcommand{\matr}[1]{\mathbf{#1}}
\newcommand{\ffrac}[2]{\ensuremath{\frac{\displaystyle #1}{\displaystyle #2}}}

\usepackage[backend=bibtex,style=ieee,citestyle=numeric,date=year,mincitenames=1, maxcitenames=2,safeinputenc, url=false]{biblatex}
\graphicspath{{./figures/}}
\usepackage[linktocpage=true]{hyperref}
\hypersetup{
	colorlinks=true,
	linkcolor=blue,
	citecolor=blue,
	filecolor=magenta,      
	urlcolor=cyan,
}

\begin{document}

{\Large\noindent
\textbf{Human-in-the-loop MGA to generate energy system
design options matching stakeholder needs}
}

\noindent 
Francesco Lombardi\textsuperscript{1 *}, Stefan Pfenninger\textsuperscript{1}
\vspace{0.3cm}

\noindent \textbf{1}  TU Delft, Faculty of Technology, Policy and Management, Department of Engineering Systems and Services, Delft, Netherlands

\vspace{0.2cm}

\noindent *Correspondence: f.lombardi@tudelft.nl 

\vspace{0.3cm}

\noindent\textbf{Abstract}: 
The common use of cost minimisation to support energy system design decisions hides from view many economically comparable design options that stakeholders may prefer. Modelling to generate alternatives (MGA) is increasingly popular as a way to go beyond least-cost designs, providing stakeholders with diverse portfolios to appraise. However, generating all the feasible designs is not computationally viable; modellers must choose what design features to generate diversity around, despite not knowing which trade-offs matter the most in practice. Therefore, MGA alone cannot ensure the generation of design options that match stakeholder needs. To address this shortcoming, we propose a human-in-the-loop (HITL) approach that automatically integrates stakeholder preferences into MGA. We elicit preferences by letting stakeholders interact with a tentative MGA design space. Hence, we decode those preferences to feed them back to the MGA algorithm and perform a guided search. This search produces a human-trained design space with more designs that mirror the elicited preferences. A synthetic experiment for the Portuguese energy system shows that HITL-MGA may facilitate consensus formation, promising to accelerate technically and socially feasible energy transition decisions.

\vspace{0.3cm}

\vspace{0.3cm}

\pagebreak

\begin{refsection}

\section{Introduction} \label{intro}

Despite the urgency of climate mitigation, progress on the energy transition remains insufficient \cite{iea_breakthrough_2023}. One of the key reasons for such insufficient progress is the difficulty in finding carbon-neutral energy system designs that are both technically feasible and acceptable for the many involved stakeholders, from citizens to industry, system operators, and policymakers \cite{susser_why_2022}. Energy system optimisation models are increasingly looked at as a way to support these complex planning decisions, but severe limitations hamper their effectiveness. In particular, previous work has shown that the conventional practice of optimising the system design for cost hides from view many other feasible designs that are economically comparable -- i.e., they are marginally more costly, or within the cost uncertainty range -- while potentially having other un-modelled benefits that make them substantially more practically viable \cite{lombardi_policy_2020}.

Approaches known as `Modelling to Generate Alternatives', or MGA, first developed in the late 70s \cite{brill_use_1979} make it possible to go beyond few, cost-optimal solutions and enlarge the decision space to hundreds of different possibilities, giving room for stakeholders to discuss and identify a consensus solution \cite{decarolis_using_2011,decarolis_modelling_2016}. They have thus seen a spike in popularity: in the last few years, novel versions of MGA have been proposed for the specific needs of state-of-the-art large-scale energy system optimisation models that, representing large geographical regions at a high spatial resolution, entail countless degrees of spatial and technological freedom to achieve a feasible system design. Such novel MGA versions aim, for instance, at generating distinct spatial configurations of infrastructure deployment \cite{lombardi_policy_2020} or at providing more homogeneous samples of the very large space of feasible decisions \cite{pedersen_modeling_2021,neumann_broad_2023}. Increasingly, MGA is recognised as a critical tool to deliver meaningful, more transparent results \cite{pickering_diversity_2022}; however, there are computational trade-offs in the use of MGA. For models of very large size, it is not yet computationally affordable to generate all the relevant design alternatives, and MGA algorithms need to decide what kind of diversity to prioritise in the design space. For instance, they must decide whether to focus more on discovering alternative technology mixes or different spatial configurations of technology deployment, for a given available amount of computational power \cite{lombardi_what_2023}. And yet, determining which design options are more worth focussing on, insofar as they may be more relevant than others in facilitating real-world consensus formation, is not trivial and should not be dealt with by modellers alone to avoid biases \cite{baker_just_2022}.

From this perspective, stakeholder engagement has untapped potential as a way to ensure that the customisation of the MGA algorithm's search strategy goes in the direction required by real-world information needs \cite{lombardi_what_2023}. A stakeholder-informed MGA search could enable a targeted, efficient use of limited computational power while simultaneously answering the increasing calls for tighter integration between scientific and practical knowledge in the co-creation of model results \cite{mcgookin_advancing_2024}. \citeauthor{finke_modelling_2024} \cite{finke_modelling_2024} highlight how stakeholder knowledge could be beneficial at various stages of an MGA workflow, for instance, informing what design features to prioritise and what boundaries to set for such features in the exploration of alternatives. They advocate for future work to investigate explicitly how to implement a similar exchange of knowledge more systematically. Similarly, \citeauthor{rosenberg_blended_2015} \cite{rosenberg_blended_2015} envisions that fast MGA workflows, enabled by ex-post sampling techniques, may facilitate the refinement of the design space via iterative interaction between modellers and stakeholders in a live setting. While acknowledging that such an interaction is paramount, they do not provide systematic methods to implement it. A recent pre-print by \citeauthor{lau_mgca_2024_preprint} \cite{lau_mgca_2024_preprint} follows in the same footsteps, leveraging dimensionality reduction and metric interpolation techniques to make MGA more computationally efficient, and thereby possible to refine in a live setting based on stakeholder feedback. The underlying assumption is that stakeholder knowledge and preferences are critical to guiding MGA towards meaningful design options, but it remains unclear how such information could be systematically streamlined, particularly in a multi-stakeholder setting. Overall, although the potential for consensus formation brought about by stakeholder integration is increasingly clear, it is yet to be understood how exactly modellers may effectively incorporate multiple stakeholders' knowledge of the problem into an MGA-based modelling workflow. 

To address this gap, we develop a first-of-its-kind human-in-the-loop (HITL) approach that seamlessly and automatically integrates stakeholder preferences into an MGA-based workflow. We propose eliciting stakeholders' system design preferences through a first interaction between stakeholders themselves and an initial, tentative MGA-based design space. Then, we feed such preferences back to the MGA algorithm to perform a `guided search' \cite{meignan_review_2015}, which produces an updated design space with a higher share of designs that match the elicited stakeholder preferences. As the new design space stems from integrating human preferences and knowledge of the problem in the automated search, we can call it a `human-trained design space'. We hypothesise that this design space facilitates the identification of a consensus solution that compromises between the many and possibly conflicting stakeholder preferences.

To test the above hypothesis, we perform an experiment under synthetic conditions in a case study of the Portuguese energy system. For our experiment, we rely on the SPORES MGA algorithm \cite{lombardi_policy_2020}, which lends itself to the proposed guided search due to its high degree of customisability, as we highlighted in previous work \cite{lombardi_what_2023}. The synthetic conditions, represented by our use of synthetic stakeholder preferences, allow us to have a clear benchmark to quantitatively assess the relative merits of the updated, human-trained design space in terms of its degree of alignment with the assumed preferences and its capacity to facilitate consensus formation. Our results substantiate that the novel HITL-MGA approach comes with the expected advantages, and shows promise for generalisability and real-world application as a means to support the acceleration of technically and socially feasible energy planning decisions.

\section{Methods} \label{methods}

In the following sub-sections we illustrate in greater detail the proposed HITL-MGA workflow and the experimental set up by which we test its effectiveness (sub-section \ref{overall_workflow}). This includes a discussion of the specific version of the SPORES MGA algorithm we use (\ref{SPORES}), and the original automated method we devise for translating human preferences into a guided SPORES search (\ref{decoding_and_translation}). Second, we discuss the energy system model of Portugal we employ for testing the workflow (sub-section \ref{model_portugal}). 

\subsection{HITL-MGA workflow and synthetic experiment setup} \label{overall_workflow}

\begin{figure}[H]
    \noindent\makebox[\textwidth]{\includegraphics[width=1.2\textwidth]{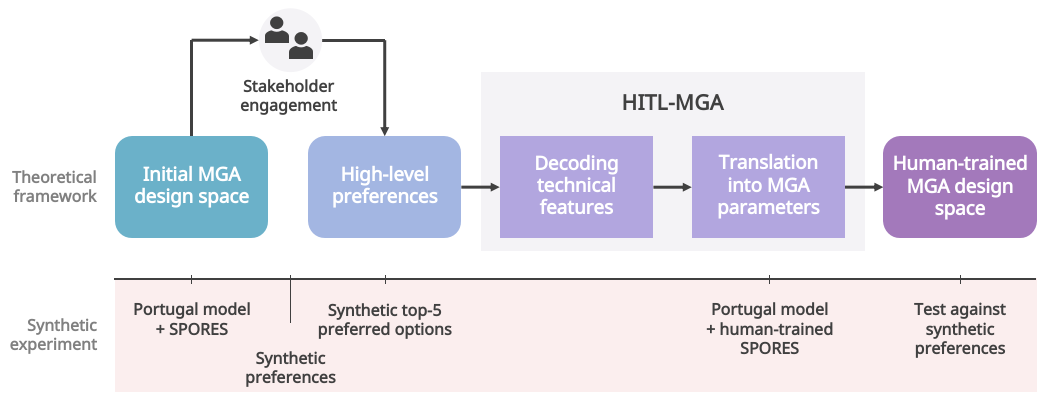}}
    \caption{Conceptual representation of the proposed HITL-MGA workflow and of the synthetic experiment we carry out in this study to test the workflow. The theoretical framework we propose envisions that we collect the high-level system design preferences arising from a first iteration of an MGA design space with stakeholders. Then, we decode which technical features underlying such preferences and translate them into parameters that may inform and re-tune the MGA search for design options. The resulting updated design space is richer in design options that align with stakeholders' high-level preferences, and it facilitates consensus formation. In our synthetic experiment, we replace stakeholder interaction with synthetic high-level stakeholder preferences, from which we derive five top-rated designs to initiate the HITL-MGA workflow. This way, we can test the outputs of the HITL-MGA workflow against the initial high-level preferences in a way that would not be possible if interacting with real-world stakeholders whose high-level preferences are typically unknown.}
    \label{fig:method_workflow}
\end{figure}

The idea of the proposed HITL-MGA workflow is to refine the system design space generated via MGA by leveraging human preferences and knowledge of the problem, which get translated into machine-readable parameters. The steps to implement the idea, summarised in Figure \ref{fig:method_workflow}, are the following. 

First, we generate an initial, best-guess MGA system design space\footnote{We provide some discussion on what MGA formulations might work best for an initial exploration of the design space in previous work \cite{lombardi_what_2023} }. Second, we let stakeholders explore this initial design space, for instance, via a web interface similar to what we proposed in previous work \cite{pickering_diversity_2022}, and ask them to select their favourite system design options. For example, stakeholders may prefer systems with a lower import dependency or perceived as more decentralised. We call these `high-level' preferences because they do not necessarily relate to concrete, particular features of the system but can be more abstract or concerned with the macro system.  For our method to work, it is not necessary to understand the rationale behind the expressed preferences; the method only requires a list of the most-selected designs. From this, we use automated decoding, which we discuss further in sub-section \ref{decoding_and_translation}, to determine the technical features inherent to a given `high-level' preference. For example, the decoding may allow us to recognise that substantial amounts of offshore wind power combined with significant domestic power-to-hydrogen capacity and little expansion of cross-border electricity transmission lines are the key features of a system design highly favoured by stakeholders. This might represent the high-level preference for "low import dependency". Unlike the high-level preference, the technical features it can be decomposed into can be translated into parameters for a customised MGA search that favours or penalises the detected features while simultaneously generating different ways of doing so. The resulting customised MGA search gives rise to an updated system design space, now richer in designs that fulfil stakeholders' preferences to varying degrees and by means of a diverse range of spatial and technological system configurations.

In this study's synthetic experiment, we test the method based on synthetic high-level stakeholder preferences, which allows us to benchmark the outputs in a way that would not be possible if interacting with real-world stakeholders whose high-level preferences, as anticipated above, may be unknown. We assume five high-level preferences:

\begin{enumerate}[noitemsep,nolistsep]
    \item A low dependency of the system on energy imports from abroad.
    \item A limited concentration of onshore wind farms in any given region.
    \item A limited overall deployment of new infrastructure to achieve the energy transition.
    \item A limited reliance on large-scale infrastructure that requires central planning, such as onshore wind farms, and thus, an increased possibility for bottom-up community initiatives to lead the transition.
    \item A low reliance on hydrogen.
\end{enumerate}

We consider these five high-level preferences to be plausible based on the insights from our SEEDS project\footnote{\href{https://seeds-project.org}{seeds-project.org}}, which included interviews with stakeholders in Portugal on this matter. We filter the initial design space generated through the combination of an energy system model of Portugal (\ref{model_portugal}) and our MGA SPORES algorithm (\ref{SPORES}) to alternatively subset it to only those system designs that perform within the top 10\% of all designs on each of these high-level features. Table \ref{tab:preference_metrics} provides further details on how we quantitatively assess a given design's performance with respect to each preference. Within each of these top-10\% subsets of designs, we randomly select one option as the representative most preferred or top-rated design for a given high-level preference, ultimately leading to a set of five synthetic top-rated designs. In a real-world case, we would have such a selection of designs based on stakeholder ranking, without knowing what high-level preferences the designs represent. We feed the set of designs into the HITL-MGA workflow outlined above, which produces an updated design space for Portugal. Finally, we test this new design space against the initial synthetic high-level preferences, assessing whether there are more or fewer design options that match those preferences. In addition, as a proxy for the likelihood of consensus formation, we quantify how many design options exist in the neighbourhood of the design that ensures the best compromise between all preferences. 

To quantify the above capacity of a system design to compromise between the different high-level preferences, we use a simple Multi-Crtieria Decision Analysis (MCDA) approach. In particular, we rely on a typical weighted-sum model \cite{fishburn_letter_1967}, in which we assess the performance of the given system design for each of the five assumed synthetic high-level stakeholder preferences, then we sum these performances together with equal weights, as per Equation \ref{eq:mcda}.

\begin{equation} \label{eq:mcda}
    {A_{q}}^{\operatorname{MCDA-score}} = \sum\limits_{k=0}^n{{w_{k}}a_{qk}}
\end{equation}

where ${A_{q}}^{\operatorname{MCDA-score}}$ is thus the total score obtained by a given alternative design option ${A_{q}}$ when considering all high-level stakeholder preferences simultaneously ($n=5$); this corresponds to our proxy for the likelihood of consensus formation of the given design. $a_{qk}$ is the performance of the design ${A_{q}}$ in terms of the high-level preference $k$, and ${w_{k}}$ is the weight of preference $k$ compared to the others. In our case, we assume all high-level preferences have equal weight.

\subsubsection{SPORES MGA algorithm} \label{SPORES}

For this study, we use a specific MGA algorithm, SPORES, which we presented in earlier work \cite{lombardi_policy_2020} and then continuously developed \cite{pickering_diversity_2022,lombardi_what_2023}. The key feature of SPORES is that it targets both technological and spatial diversity of system designs, under the assumption that not only \textit{what} technologies are deployed but also, and in particular, \textit{where} they are deployed is critical for real-world consensus formation. What is more, compared to other MGA algorithms, SPORES is highly customisable \cite{lombardi_what_2023} and thus lends itself to being adapted for the guided search informed by stakeholder preferences. 

\begin{figure}[H]
    \noindent\makebox[\textwidth]{\includegraphics[width=1\textwidth]{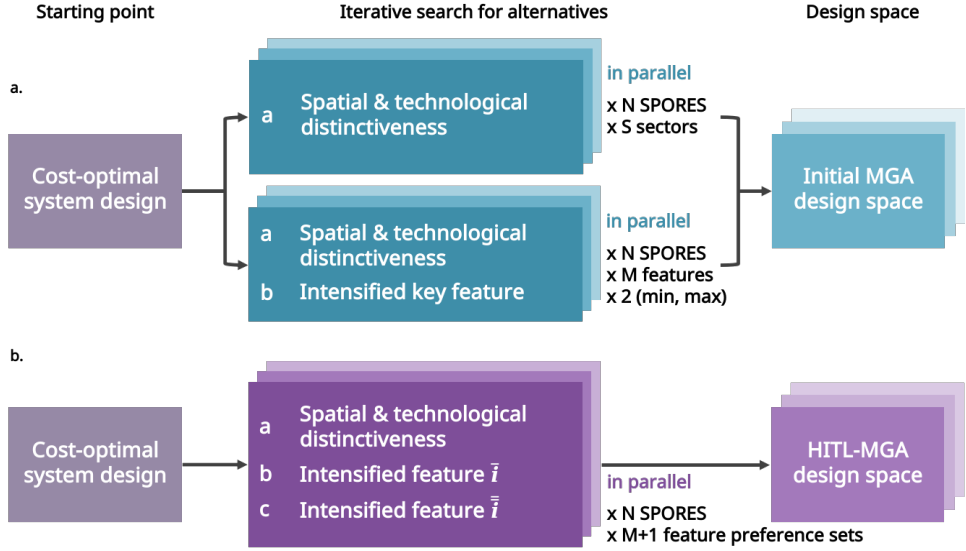}}
    \caption{Stylised representation of the SPORES algorithm, in its standard (\textbf{a}) and HITL (\textbf{b}) formulations. Starting from the cost-optimal solution, the algorithm iteratively looks for additional feasible and only marginally costlier system designs in parallel batches. The standard formulation (\textbf{a}) foresees S parallel batches that generate alternative designs based on the sole objective of making them spatially and technologically distinct from previously found solutions. Each of these S batches targets decision variables pertaining to a specific energy sector, such as power, heating, mobility or synthetic fuels. At the same time, 2xM additional batches generate (in parallel) alternatives around the intensification of specific features of the design space, for instance, the very high or very low deployment of offshore wind power overall. In the HITL formulation (\textbf{b}), each parallel batch is informed by a high-level stakeholder preference M and intensifies one or more technical features that underlie such a preference while also targeting spatial and technological distinctiveness from previously found solutions. One additional batch (M+1) targets all the given high-level preferences simultaneously.}
    \label{fig:spores_concept}
\end{figure}

Figure \ref{fig:spores_concept}a illustrates the high-level implementation of the algorithm, which envisions one default run with a single MGA objective and several parallel runs with a multi-objective MGA formulation aimed at generating alternatives from different corners of the near-optimal solution space. The basic idea underpinning this formulation is the following. We identify the cost-optimal solution as a starting point. Then, the algorithm searches for new feasible system designs that are different in terms of both \textit{what} technologies are deployed and \textit{where} compared to what is found in the previous solution(s). Mathematically, we achieve this by incrementally assigning penalties, or weights, to spatially explicit capacity investment decision variables that featured prominently in previous solutions. In those runs that feature a multi-objective MGA formulation, we combine the search for spatially and technologically distinct feasible system designs with the intensification of specific design features, such as the high or low deployment of a particular technology. We define `intensification' as pushing the given feature either towards its minimum or maximum feasible value, depending on whether it is a desired or undesired feature. As illustrated elsewhere \cite{lombardi_what_2023}, this ensures that extreme technological boundaries of the design space are captured and that otherwise hard-to-discover design options are generated. In all cases, the newly generated feasible solutions are constrained to be only marginally more costly than the least-cost feasible solution (in this study, we assume this means a 10\% cost increase is acceptable, in line with previous work \cite{pickering_diversity_2022,neumann_near-optimal_2021}).

In mathematical terms, we can formulate the generic SPORES-MGA problem as in Equation \ref{eq:spores_generic}.

\begin{equation} \label{eq:spores_generic}
	\begin{array}{*{20}{c}}
		{\min\; \;{Y} = a \cdot \sum\limits_j {\sum\limits_i {{w_{ij}}x_{ij}^{cap}} \pm b \cdot \sum\limits_j{ x_{\overline {i}j }^{cap}}  } }\\
		{s.t.\;\;cos{t_n} \le (1 + s) \cdot cos{t_0}}\\
		{\matr{Ax} \le \matr{b}}\\
		{\matr{x} \ge 0,}
	\end{array}
\end{equation}

where $i$ and $j$ indicate the $i$-th technology type and the $j$-th location within the model; $x_{ij}^{cap}$ is the capacity investment decision variable for the $ij$-th location-technology pair; and $x_{\overline {i}j}$ is the capacity decision variable associated with the technology under intensification. The $a$ and $b$ coefficients are the weights associated with the different components of the objective function: when $b$ has a positive sign, the technology under intensification is minimised; for a negative sign, it is maximised. Finally, if $b$ is null, the formulation collapses into the default case with a single objective (see Figure \ref{fig:spores_concept}a). $\matr{A}$, $\matr{b}$, are a matrix and a vector of coefficients that give rise to the physical constraints alongside the decision variables' vector $\matr{x}$; $cos{t_n}$ is the total annualised system cost, which is bound to remain within a marginal relaxation of the optimal cost ($cos{t_0}$); and $s$ is the accepted cost relaxation, also known as cost slack. 

In this study, we apply SPORES to a model of the Portuguese energy system (see sub-section \ref{model_portugal}). Following the logic outlined in Figure \ref{fig:spores_concept}a, we generate a total of 80 SPORES in four parallel batches using the default single-objective mode applied to decision variables pertaining to different energy sectors (S=4): power (N=50 SPORES), heating, mobility and synthetic fuels (N=10 per sector). Then, we generate a total of 180 SPORES across multi-objective batches that alternatively maximise (N=10) or minimise (N=10) one of the following (M=9) key technology assets: wind onshore, wind offshore, wind overall, open-field solar PV, roof-mounted solar PV, biofuels, battery storage, electrolysis and transmission capacity. This leads to a total of 260 (80+180) different system design options, or SPORES, of which about 70\% adopting a multi-objective MGA formulation.

\subsubsection{Decoding of technical features and translation into MGA parameters} \label{decoding_and_translation}

As per the workflow in Figure \ref{fig:method_workflow}, once an MGA-generated design space is available, stakeholders may explore it and select their favourite design(s) based on their high-level preferences. Out of the pool of individual stakeholder preferences, we quantify the top five most frequently preferred, or `top-rated' designs. Hence, the core of the HITL-MGA approach takes place, decoding the technical features underlying these top-rated designs and translating those into MGA parameters for a new, human-trained MGA step.

In practice, for each of the top-rated designs, we compute the statistical distribution of capacity investment decision variables in key system assets across the entire decision space. The specific design and research question at stake should guide the definition of what a `key system asset' is. In our case, looking at the design options for a carbon-neutral energy system at the country scale, we consider the following key system assets: all renewable energy generation capacity, battery storage capacity, electricity transmission capacity and electrolysis capacity (as a proxy for the overall role of hydrogen and synthetic fuels in the system). Hence, we assess how these key capacity decision variables play out in the given top-rated design compared to the statistical distribution of the entire design space. Any key system asset decision variables that substantively (which we define as $\geq$ 15\%) deviate from the mean of the distribution across all initial designs are flagged as distinctive (either positively or negatively) for the design under scrutiny and, therefore, likely to contribute majorly to the high-level characteristics of the associated system that may have driven stakeholders' preference. In this study, we also perform a sensitivity analysis for a stricter (10\%) threshold in determining when a given system asset can be considered to deviate from the distribution mean substantively.

Then, we translate the identified distinctive technical features into parameters for the MGA search, leveraging the same approach used to intensify generic system features in Equation \ref{eq:spores_generic}. Positively distinct features (featuring more prominently than in the mean) are maximised, whilst negatively distinct features are minimised. Unlike in Equation \ref{eq:spores_generic}, though, we now allow for the simultaneous intensification of multiple features. The objective function changes as in Equation \ref{eq:spores_hil}.

\begin{equation} \label{eq:spores_hil}
	\begin{array}{*{20}{c}}
	{\min\; \;{Y_\mathit{HITL}} = a \cdot \sum\limits_j {\sum\limits_i {{w_{ij}}x_{ij}^{cap}} + b \cdot \sum\limits_j{ x_{\overline {i}j }^{cap}} -  c \cdot \sum\limits_j{ x_{\overline{\overline {i}}j }^{cap}}  } }\\
	\end{array}
\end{equation}

where the coefficient preceded by a positive sign ($b$) multiplies the decision variables corresponding to features to be minimised system-wide, and the coefficient preceded by a negative sign ($c$ ) refers to features to be maximised. More than one feature at a time can be maximised and minimised, which can be obtained by simply adding additional $b$ and $c$ coefficients with the corresponding signs.

We repeat the approach of decoding underlying technical features and translation into parameters for Equation \ref{eq:spores_hil} for each of the top-rated designs; in this case, five (M=5). In addition, we set up one problem in which all the preferences reflected in the five top-rated designs are taken into account simultaneously, except for those preferences that are mutually exclusive (as outlined in Figure \ref{fig:spores_concept}b, where we envision M+1 parallel runs). 

In practice, this leads to six (M=5, +1) different MGA runs that can be dealt with in parallel, similar to the original SPORES workflow. The number of designs to be generated within each of these parallel runs can be fine-tuned to match the same size of the original design space. In this study, we generate N=45 design options per parallel run, further distributing these 45 design options in three parallel batches of 15 designs each, in which we adopt different MGA weights ($w_{ij}$) to further increase diversity as done in previous work \cite{lombardi_what_2023}. More precisely, we adopt, in one case, a simple integer weighting method; in another, a method based on deployment relative to maximum potential; and in the last case, a method based on the distance from the evolving average weight (see the Supporting Methods for further details). In total, we obtain a design space of 270 system designs, slightly larger than the original one. This similarity in size between the original and the HITL-MGA option spaces is purposeful so that we can assess potential benefits for a comparable computational effort. Both the original and the HITL-MGA option space are computed on the DelftBlue high-performance computer \cite{DHPC2024}.

\subsection{Model of the Portuguese energy system} \label{model_portugal}

The model used to generate the results (named ‘Calliope-Portugal’) uses the open-source Calliope modelling framework \cite{pfenninger_calliope_2018} and on the model data for all European countries available as part of the Euro-Calliope project. In particular, Calliope-Portugal inherits the full characterisation of energy sectors (electricity, heat, transport and industry) from the Sector-Coupled Euro-Calliope model \cite{pickering_pre-built_2021}, but implements (exclusively for Portugal) the higher spatial granularity for wind and solar available from the electricity-only Euro-Calliope model \cite{trondle_euro-calliope_2020}. 

The model accounts for Portugal's entire energy demand across all sectors, spatially distributed between two macro-nodes, North and South. The model's task is to design the system in such a way as to satisfy the demand for a full year of operation based on carbon-neutral energy technologies, which broadly consist of renewables, sustainable biofuels, and green hydrogen and derived e-fuels, whose costs take into account projections towards 2050. The model only considers the final system design, or a `snapshot' of what the system could look like in 2050, disregarding capacity deployment pathways through time on the way to such a final design. The two macro-nodes of North and South are connected via electricity transmission based on the network topology available from the Sector-Coupled Euro-Calliope. Within each node, however, renewable power capacity can be deployed in several sub-regions (18 in total), corresponding to the country's administrative regions, with different land availability and capacity factors. Countries outside of Portugal are eliminated from the model and substituted with a simple representation of import and export options for electricity and hydrogen at a fixed price. For electricity, the price is set based on the historical average electricity price for trade with neighbouring Spain. For hydrogen, the price per kWh is calculated assuming a market price of 1.5 EUR/kg for green hydrogen in 2050 and 33.3 kWh/kg as the lower heating value. In this study, we use 2016 as the reference weather year, since we quantified it as a `typical' year for Portugal within those available from the Euro-Calliope dataset.

All the other model assumptions and their underlying rationale are available from the original publication that featured the Sector-Coupled Euro-Calliope model \cite{pickering_diversity_2022}, to which we refer the reader for additional information. The Calliope-Portugal model files are publicly available on Zenodo \cite{lombardi_calliope-portugal_zenodo}. 

\section{Results and discussion} \label{results}

To assess the proposed HITL-MGA workflow, we first look at whether the human-trained system design space is richer in designs that align with the assumed synthetic high-level stakeholder preferences. We do so for the reference case of a 15\% deviation from the mean as a threshold for feature distinctiveness and provide a sensitivity analysis for the case of a narrower (10\%) threshold. Second, we analyse those designs that best ensure a compromise between all stakeholder preferences, for both the original and the human-trained design spaces. In doing so, we look for illustrative design options that are available in the human-trained design space and that may be particularly attractive to stakeholders and that are not available in the original design space.

\subsection{The human-trained designs better align with high-level preferences in most cases} \label{preference_matching}

\begin{figure}[H]
    \noindent\makebox[\textwidth]{\includegraphics[width=1.3\textwidth]{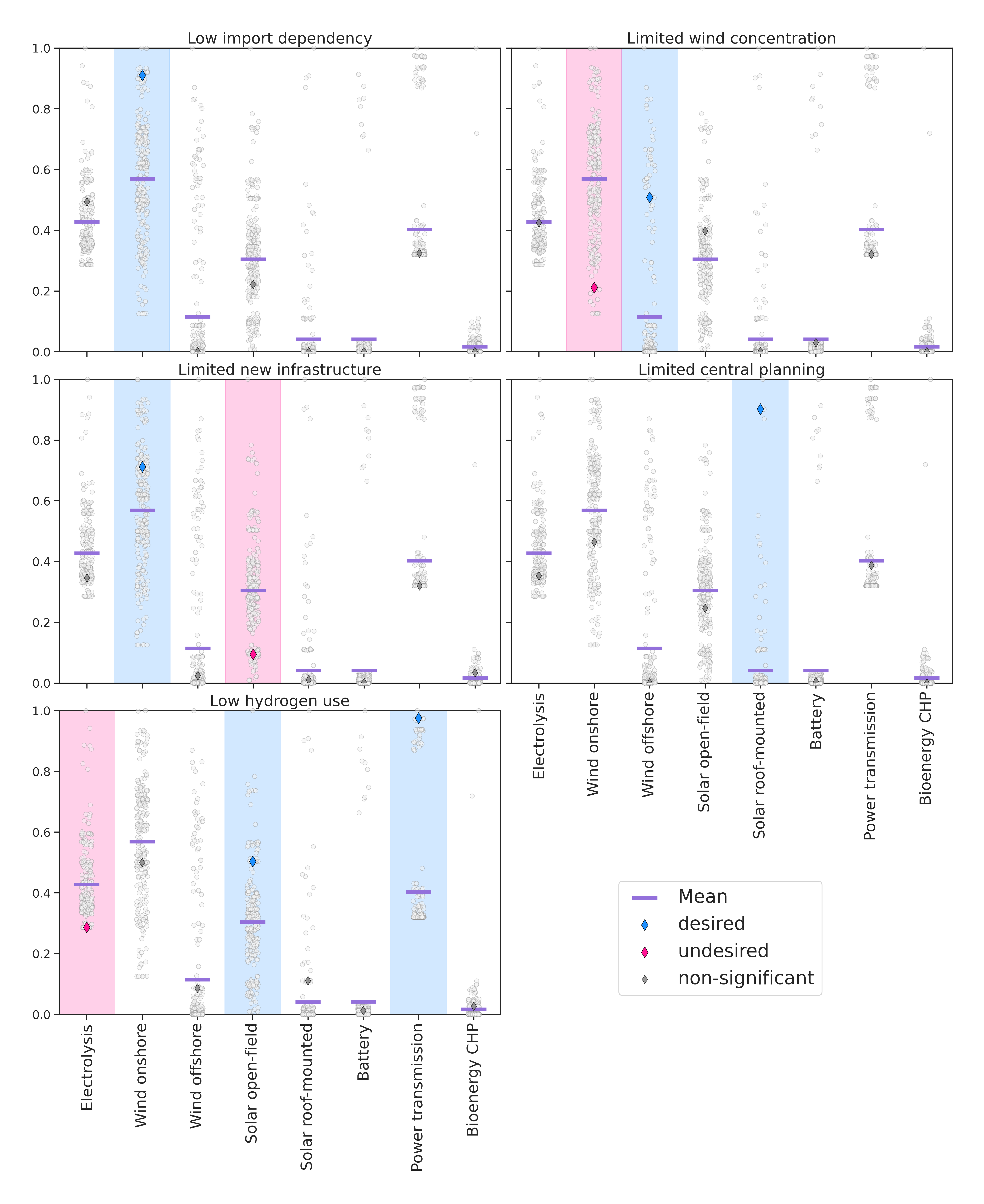}}
    \caption{Results of the automated decoding of distinctive technical features in each top-rated system design based on comparison with the statistical distribution (represented by strip plots) of technical system features across the entire design space. In the figure, the reliance on each technical feature is normalised with respect to the maximum deployment of the given feature experienced across the entire design space. ‘Desired’ features are highlighted in blue, whilst ‘undesired’ ones are highlighted in red. In this case, the threshold for detecting a given feature as desired or undesired is set to a 15\% deviation from the distribution's mean before normalisation.}
    \label{fig:decoding_multipanel_15}
\end{figure}

As outlined in sub-section \ref{decoding_and_translation}, our HITL-MGA workflow is grounded in the automated decoding of the distinctive technical features underlying each of the top five system designs preferred by stakeholders in their initial appraisal of the solution space. Figure \ref{fig:decoding_multipanel_15} illustrates the results of the decoding step. For instance, for the system design that we chose to represent the high-level preference `low regional concentration of onshore wind farms'. Using the statistical decoding logic discussed in sub-section \ref{decoding_and_translation}, the overall deployment of onshore wind power is detected as an undesired feature, whilst the deployment of wind offshore is detected as desired. Figure \ref{fig:decoding_multipanel_10} provides the outcome of the procedure for each of the other top five preferred designs that we chose as representatives of their respective high-level preference. These results illustrate how, even for those high-level preferences that have a less abstract nature, the decoding procedure may provide additional nuances to guide the MGA search. For instance, for the highly-rated design embodying the preference for a low reliance on hydrogen, the decoding detects the deployment of electrolysers as undesired, but alongside a high deployment of solar PV farms and an expansion of the electricity grid -- complementary features that would have been difficult to guess based on the high-level preference alone.

\begin{figure}[H]
    \noindent\makebox[\textwidth]{\includegraphics[width=1.2\textwidth]{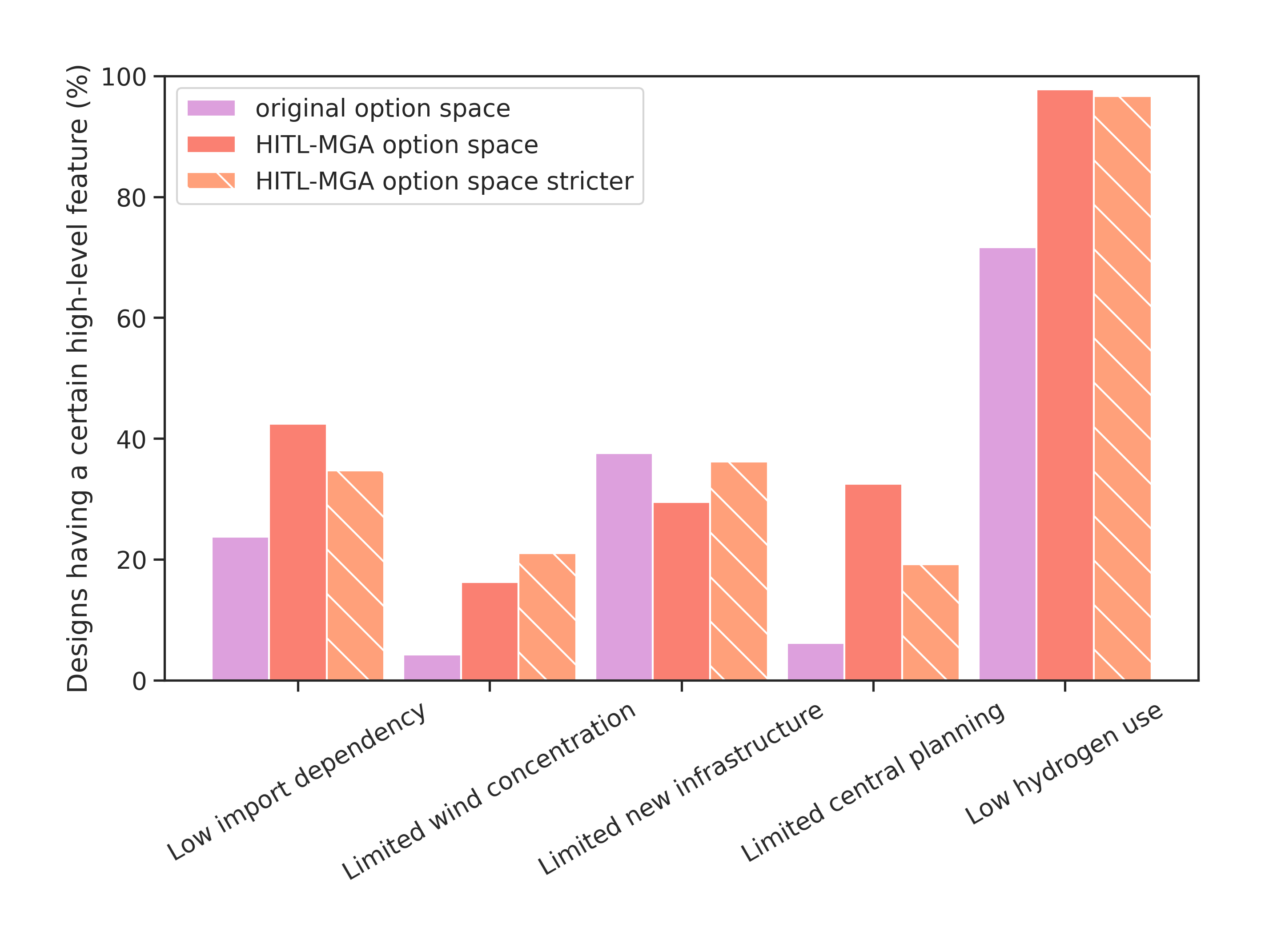}}
    \vspace{-1cm}
    \caption{Comparison of the original design space generated via standard MGA and the one generated with the HITL-MGA workflow in terms of percentage of designs matching the assumed high-level stakeholder preferences. HITL-MGA results are shown both for the default case of a 15\% threshold in the decoding procedure and for the case of a stricter (10\%) threshold.}
    \label{fig:high_level_matching}
\end{figure}

As anticipated in section \ref{intro}, the new design space incorporates human preferences and knowledge of the problem, so we refer to it as the `human-trained design space'. It results from feeding the SPORES MGA algorithm with the decoded technical features that represent a high-level preference, plus all the (non-mutually-exclusive) features decoded across all preferences simultaneously. This produces a richer set of designs that align with the given high-level preferences, as can be seen in the comparison shown in Figure \ref{fig:high_level_matching}. In particular, the human-trained design space has substantially more designs matching those high-level preferences with a more direct connection to the deployment of specific technologies, such as the low use of hydrogen or the limited reliance on large-scale infrastructure that requires central planning. More nuanced preferences, such as the low regional concentration of wind farms, do improve, but less markedly. Finally, the high-level preference for a system design entailing a limited overall deployment of new infrastructure, represents an exception for which the human-trained design space falls short in increasing the number of design options. This is a consequence of the method's design, which relies on the intensification of features (Equation \ref{eq:spores_hil}) to guide the search. These features are the capacities of technologies to be deployed.  Thus, pushing technology capacities to their low or high extremes, especially when not diluted by doing so for multiple features at once, easily leads to more solutions with an overall high deployment of capacity, compared to the original design space that also entailed 80 solutions generated without any intensification (see sub-section \ref{SPORES}). 

One option to increase the number of features decoded as distinctive and hence intensified (i.e. minimised or maximised) simultaneously, is to reduce the distance-from-the-mean threshold for decoding. Figure \ref{fig:decoding_multipanel_10} shows the outcome of the decoding procedure for a reduced (from 15\% to 10\%) threshold, which results in the design space labelled as `stricter' in Figure \ref{fig:high_level_matching}. As expected, in this case, the percentage of designs with limited overall deployment of new infrastructure improves compared to Figure \ref{fig:high_level_matching}, although it is still marginally less than in the original design space. Moreover, the percentage of designs aligned with a preference for limited central planning decreases, due to the `noise' introduced by the additional features to be minimised or maximised alongside the push for less-centralised technologies, such as rooftop solar PV. There is hence a trade-off in having a stricter or more relaxed threshold for the decoding of distinctive technical features. 

Regardless of the chosen threshold, the increase in design options that align with high-level preferences across most categories far exceeds the marginal decrease experienced in a single category, which suggests that the potential for consensus formation likely increases as well. We now dive deeper into this aspect.

\subsection{The potential for consensus formation increases substantially} \label{consensus_results}

Having assessed the overall increase in system designs that align with specific high-level preferences, we move on to quantifying to what extent the human-trained design space may improve the likelihood of consensus formation. As anticipated in section \ref{overall_workflow}, as a proxy for this likelihood, we use the overall score obtained by a given system design when applying a multi-criteria decision analysis (MCDA) to the five synthetic high-level preferences, which are weighted equally. The more a design performs well across all the five assumed high-level preferences, the higher its MCDA score and its potential to represent a compromise between conflicting preferences. After applying the MCDA and obtaining an aggregate score for all designs, we first select the overall best-performing design across both the MGA and HITL-MGA design spaces. In our case, this design happens to lie in the human-trained (HITL-MGA) space. We then analyse which and how many designs exist that perform nearly as well, which we define to be no as scoring now less than 25\% below the overall best design. We call these design options the `near-highest-consensus' designs.

\begin{figure}[H]
    \noindent\makebox[\textwidth]{\includegraphics[width=1.1\textwidth]{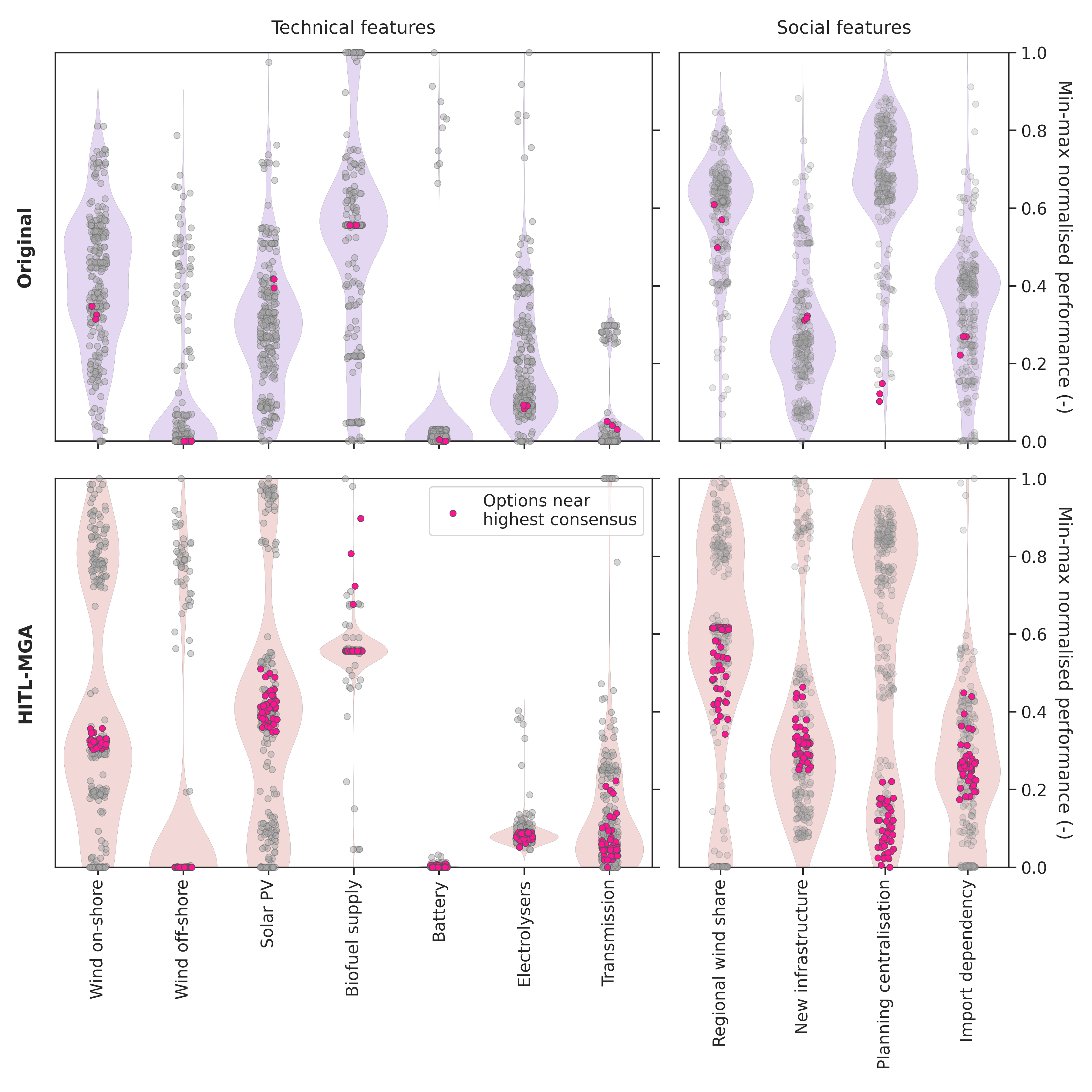}}
    \vspace{-0.5cm}
    \caption{Solution space generated by the original simple MGA and the HITL-MGA workflows from the perspective of technical (system-wide capacity deployment of a given technology) and social system features of stakeholder interest. We show the designs (x-axis) by their performance on a given metric compared to the maximum experienced across both design spaces for the same metric (y-axis). To ensure comparability, results for each metric are re-scaled between 0 and 1 based on the minimum and maximum values experienced across either design space (an approach also known as min-max normalisation). In each design space, the designs highlighted in red are those that we define as `near-highest-consensus'. They are within 25\% of the system design option that, across both design spaces, ensures the best performance across all synthetic high-level preferences (see sub-section \ref{overall_workflow}).}
    \label{fig:option_spaces_result}
\end{figure}

Figure \ref{fig:option_spaces_result} shows the shape of both the original and the human-trained design spaces alongside the results of the MCDA analysis. The near-highest-consensus designs are much more abundant in the human-trained design space, about 18\% of whose designs are close to the best performing one, compared to just about 1\% in the original design space. In fact, while the original design space provides a more homogeneous distribution of design options in some categories -- for instance, the deployment of electrolysers --, by its very layout, the human-trained design space creates an abundance of design options precisely in those areas of the feasible design space towards which human inputs guided it -- such as the design space characterised by a low deployment of electrolysers. As a result, while the updated design space is apparently less diverse overall, it offers more degrees of freedom around features of interest, exactly as envisioned with the conceptualisation of the algorithm. In addition, this targeted search for design features underlying certain high-level preferences is complemented by a search that targets simultaneously all the features decoded across all high-level preferences (as synthesised by Figure \ref{fig:spores_concept} and discussed in sub-section \ref{decoding_and_translation}). This additional search was envisioned to generate in-between options that could facilitate consensus formation. Figure \ref{fig:option_spaces_result_per_run} confirms the effectiveness of this algorithm design choice, since the system designs generated in the above run contribute substantially to the near-highest-consensus subset.

\begin{figure}[H]
    \noindent\makebox[\textwidth]{\includegraphics[width=1.2\textwidth]{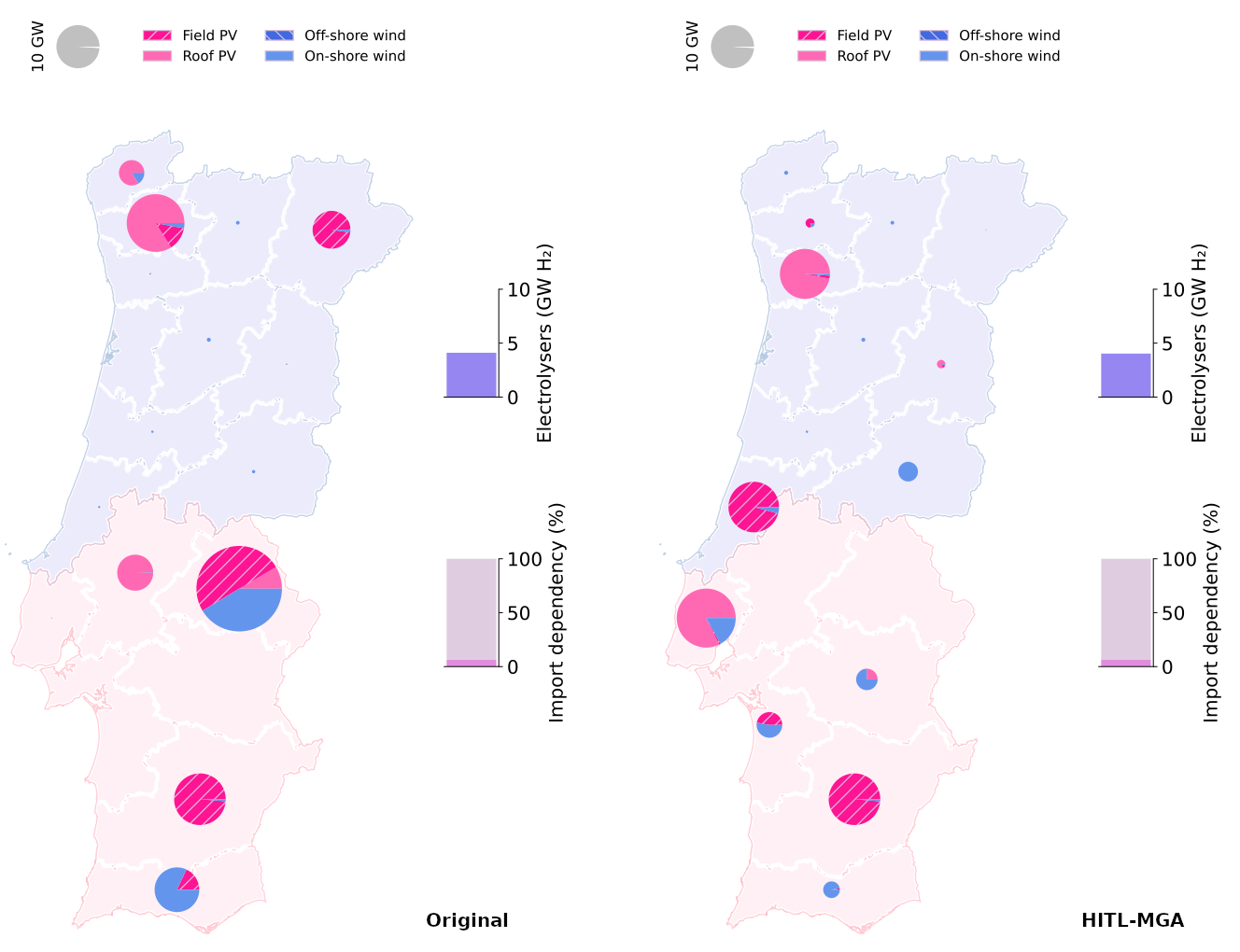}}
    \caption{Spatial deployment of renewable energy generation capacity, system-wide deployment of electrolysers and degree of reliance on imports in the system's primary energy supply. The results are shown for two illustrative system design options, namely the options -- in the original and in the human-trained (HITL-MGA) option spaces -- with the best performance in terms of limited regional concentration of wind farms among the subset of overall best-performing, or `near-highest-consensus', options. The geodata source to produce the maps is GISCO-Eurostat \cite{eurostat_nuts}. }
    \label{fig:illustrative_maps}
\end{figure}

When it comes to facilitating consensus formation, the human-trained design space thus provides several designs that allow achieving a good compromise between all the five high-level preferences outlined in sub-section \ref{overall_workflow}, and that still do so in different ways, leaving room for further discussion. In practice, this means that the updated design space opens up possibilities for real-world discussion that were missed by the initial MGA exploration. As an illustrative example, we may look at the case of the share of wind farms in any given region that, in our experimental setup, stakeholders prefer to limit. The human-trained design space provides a substantial degree of flexibility with respect to wind farms concentration even within the subset of near-highest-consensus designs (Figure \ref{fig:option_spaces_result}). If we select, within such a subset and for each design space, the design that provides the best performance in terms of limited concentration of wind farms, its spatial configuration of infrastructure deployment looks strikingly different in the two cases, as shown in Figure \ref{fig:illustrative_maps}.

In the case of the original design space (Figure \ref{fig:illustrative_maps}a), two regions host most of the onshore wind power capacity. In the system design selected from the human-trained design space for the same criterion (Figure \ref{fig:illustrative_maps}b), instead, onshore wind capacity is much more fairly distributed across regions, even though the overall capacity is nearly identical (\textasciitilde 20 GW) compared to the previous case. At the same time, the system design selected from the human-trained design space has a comparable (marginally better) performance in terms of electrolysers deployment, total deployment of new infrastructure (a proxy for the rate of transition) and import dependency. The share of decentralised technologies, such as rooftop solar PV, which may limit central planning in favour of bottom-up, citizen-led initiatives is also comparable between the two designs. Thus, the design option made available by the human-trained design space substantially outperforms any option available in the original design space in terms of limiting the regional concentration and the total amount of onshore wind power capacity, while simultaneously ensuring an equally good performance across all other high-level stakeholder preferences. This example illustrates how consensus formation can be facilitated by the HITL-MGA approach.

\section*{Conclusion}

With this work, we set out to investigate whether an innovative human-in-the-loop approach to MGA could generate energy system design options that better align with stakeholder preferences and facilitate consensus formation compared to conventional, modeller-driven MGA. The results obtained by our original HITL-MGA workflow for a case study of the Portuguese energy system under synthetic conditions (i.e., synthetic stakeholder preferences) show that this is the case, while also highlighting areas for further development.

In particular, our results highlight that setting up a HITL-MGA workflow can be achieved by combining two key ingredients. First, a statistics-based automated decoding of the technical features underlying given high-level stakeholder preferences. Second, a recalibration of the MGA objective function's parameters to perform a guided search. While we demonstrated the MGA recalibration for the specific case of the SPORES MGA algorithm, the parameters we manipulate in SPORES are common to most other recent MGA approaches applied to energy models literature \cite{neumann_broad_2023,patankar_land_2023,sasse_low-carbon_2023} and are otherwise trivial to include in the objective function.

Moreover, we show that our HITL-MGA workflow effectively generates more designs that match the initial hidden high-level stakeholder preferences. Nonetheless, as an exception, we also detect that the intensification of features envisioned by our HITL-MGA workflow is structurally bound to conflict with potential high-level preferences for limited new infrastructure deployment. In other words, intensifying a given technical feature comes with a substantial deployment of the related infrastructure. In our experiment for the case of Portugal, the only high-level preference for which the human-trained design space does not produce an improvement is the preference for a low rate of infrastructure deployment, which in our model setup coincides with less deployment overall. This is not surprising, since the nature of the energy system model used here is to model the deployment of technologies, and we use intensification (maximisation or minimisation) of technology capacity as the way to decode stakeholder preferences. Future work should examine the use of simple additional statistical measures to detect a preference for limited new infrastructure and ways to translate that into the guided search. A possible translation could be reducing the strength of the \textit{b} and \textit{c} coefficients in Equation \ref{eq:spores_hil}, representing the intensification of desired and undesired features, relative to coefficient \textit{a}, aiming at the diversification of the system design. In previous work, we have already shown that such a manipulation effectively mitigates the degree of intensification of technical features \cite{lombardi_what_2023}.

We also observe that the need to define a threshold for the automated decoding of statistically distinctive technical features in a given top-rated design entails a certain degree of uncertainty in the outcome of the decoding and, in turn, the human-trained design space. A tighter threshold, which we also examined, results in the decoding step detecting more features as distinctive, spreading and diluting the intensification objectives in Equation \ref{eq:spores_hil} across more features. This may work in favour of some high-level preferences, such as the one for limited infrastructure deployment overall, but less so for others. In general, the exact value of the threshold cannot be easily connected to any objective measure and, therefore, should be tested and calibrated beforehand, for instance, by means of a synthetic experiment like the one we carry out in this study, in such a way as to match the needs of the model and research questions at stake.

Regarding the likelihood of consensus formation, which we approximate via a simple MCDA analysis aligned with the assumed stakeholder preferences, the HITL-MGA workflow largely outperforms conventional MGA. In particular, the share of designs with an MCDA score within 25\% of the highest score achieved in either design space increases from 1\% to 18\% when human inputs are integrated into the MGA loop. This is because the major improvements recorded in the number of stakeholder-appealing designs across most high-level preferences substantially counterpoise the marginally lower number of designs matching the `low-rate of deployment' preference discussed above. In practice, the human-trained design space provides design options that allow stakeholders to compromise between their conflicting preferences, which were entirely missed by the original design space. For instance, designs that significantly limit the regional concentration and the overall deployment of onshore wind power capacity while simultaneously performing well on all the other assumed high-level preferences. Nonetheless, it is critical to underscore how, in a real-world setting, a computational approach that generates more middle-ground options by incorporating and synthesising multiple human inputs is an enabler but not a guarantee of the higher likelihood of consensus formation \cite{royston_masters_2023}. It is paramount to complement the computational side with a collaboration with experts in stakeholder engagement and participatory action research that may ensure appropriate representation of all the stakeholder groups and a constructive deliberation process \cite{mcgookin_participatory_2021,susser_rethink_2024}. 

In our synthetic experiment, we purposefully assume that the top-rated designs in the initial MGA space match some immediately understandable and not multifaceted preferences because this facilitates our testing and assessment of the improvements brought about by the HITL-MGA workflow. However, the workflow is entirely agnostic to the initial preferences and does not require any understanding of the rationale behind them. It only requires a list of the most-selected designs, which makes it applicable to any set of multifaceted real-world preferences. It is also worth noting that, while we run the HITL-MGA workflow to decode and implement back into the MGA search system-wide technical features only, high-level preferences may well refer to specific features at the local scale. For instance, some stakeholders may be against a certain technology in a certain region, such as the deployment of electrolysers in regions affected by water scarcity, despite not opposing that technology per se. Even though we have not looked into this possibility in this first synthetic experiment, nothing prevents modellers from running the same workflow for spatially explicit variables. Future work may further explore this possibility and its potential advantages. Another interesting possibility that future work may explore is the discretisation of the degree of (un)desirability of those distinctive technical features underlying a given top-rated design. Rather than simply flagging a feature as (un)desired, the method could accommodate incremental weights to further differentiate between highly or only moderately (un)desired features. Such discretisation would require a decision on how to define the discrete degrees of (un)desirability, further contributing to the need for calibration discussed earlier for the current threshold for feature distinctiveness. At the same time, it may further improve the accommodating of conflicting stakeholder preferences and the identification of middle-ground options. Finally, the HITL-MGA workflow in this study leverages the latest developments in terms of computationally efficient generation of MGA design spaces. However, the MGA field is continuously and rapidly evolving, and the proposed HITL-MGA would only benefit from any further computational advancements. Promising avenues for the further enhancement of HITL-MGA may be techniques such as the ex-post dimensionality reduction and metric interpolation proposed by a recent pre-print \cite{lau_mgca_2024_preprint} or artificial intelligence techniques that are already common in the sister field of energy system multi-objective optimisation \cite{prina_machine_2024}.

The improved consensus-formation capabilities enabled by the HITL-MGA approach, and its very nature of integrating human preferences and knowledge of the problem into a computational workflow, align with the increasing demands for a tighter cross-fertilisation between engineering and other disciplines in tackling energy transition challenges at the interface between policy and science \cite{baker_just_2022,mcgookin_advancing_2024}. Our human-trained search for design options answers the call for more `fair and unbiased algorithms' \cite{baker_just_2022} supporting energy transition decisions. And it does so by leveraging the principles of co-creation  \cite{bruckmann_towards_2023,mcgookin_advancing_2024} and providing a computational platform for their direct application.

Overall, this study is the first to propose and successfully test a human-in-the-loop approach to MGA under synthetic conditions. Future work should explore the possible improvements of the method outlined above and test it further in a `field' experiment with real-world stakeholders. The results of our synthetic experiment show that HITL-MGA is a promising way to ensure that the generation of system design options matches stakeholder needs and facilitates consensus formation, thereby enabling and accelerating a just practical implementation of the energy transition.

\section*{Acknowledgements}
This work received funding as part of the SEEDS project. The SEEDS project is supported by the CHIST-ERA grant CHIST-ERA-19-CES-004, the Swiss National Science Foundation grant number 195537, the Fundação para a Ciência e Tecnologia (FCT) grant number CHIST-ERA/0005/2019, the Spanish Agencia Estatal de Investigación with grant PCI2020-120710-2, and the Estonian Research Council grant number 4-8/20/26.

\section*{Author contributions}
Conceptualisation, F.L. and S.P.; Methodology, F.L.; Software, F.L.; Investigation, F.L.; Writing - Original draft, F.L.; Writing - Review \& Editing, F.L. and S.P.; Visualisation, F.L.;  Supervision, S.P.; Project Administration, F.L. and S.P.; Funding Acquisition, F.L. and S.P.  

\section*{Declaration of interests}
The authors declare no competing interests.
\microtypesetup{protrusion=false}
\printbibliography
\end{refsection}

\begin{refsection}
\section*{Supporting Results}

\setcounter{page}{1}
\renewcommand\thefigure{S\arabic{figure}}    
\setcounter{figure}{0}    
\renewcommand\thetable{S\arabic{table}}    
\setcounter{table}{0}   
\renewcommand\theequation{S\arabic{equation}}    
\setcounter{equation}{0}

\begin{figure}[H]
    \noindent\makebox[\textwidth]{\includegraphics[width=1.1\textwidth]{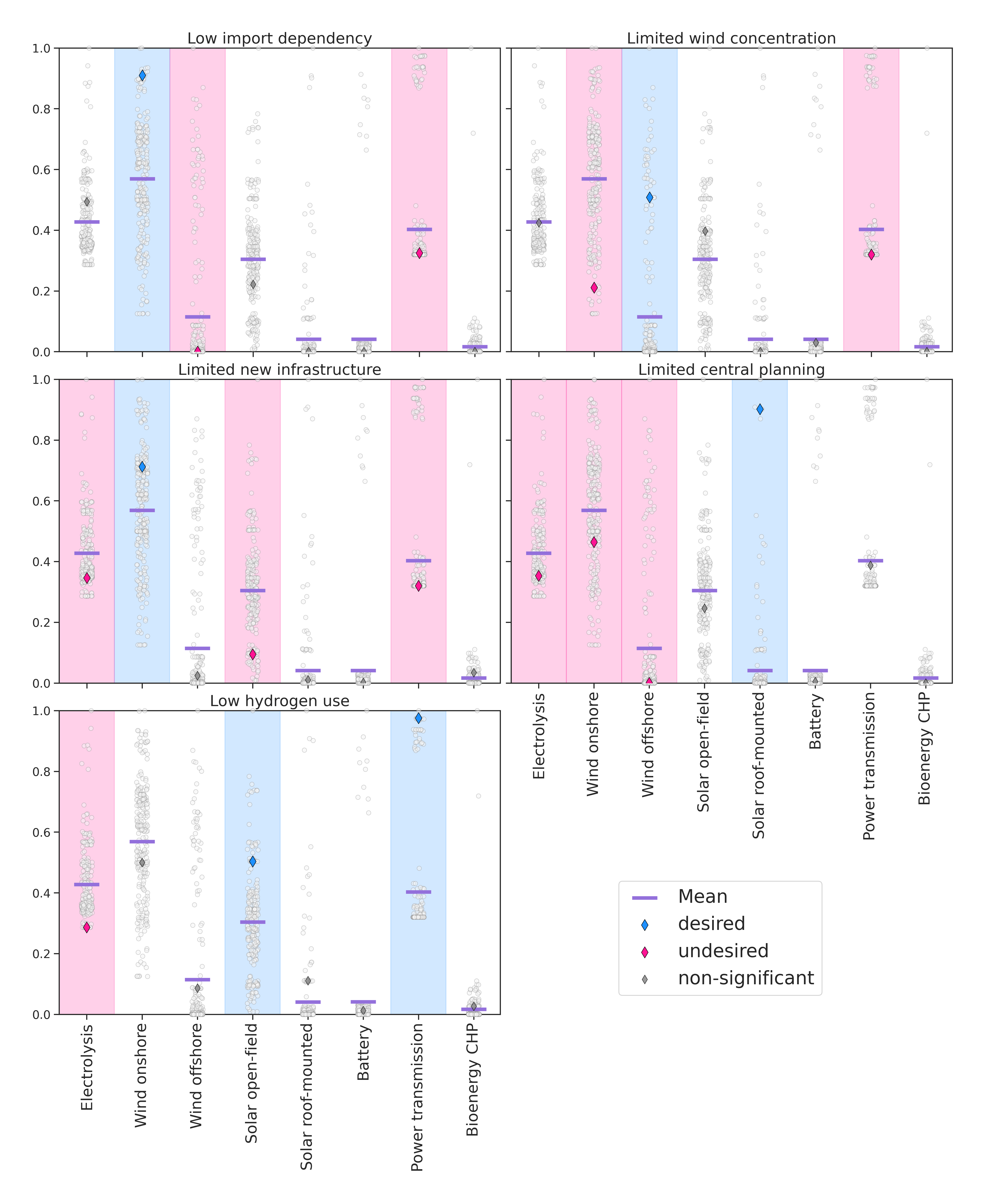}}
    \vspace{-1cm}
    \caption{Complements Figure \ref{fig:decoding_multipanel_15}. the threshold for detecting a given feature as desired or undesired is set to a 10\% deviation from the distribution's mean before normalisation.}
    \label{fig:decoding_multipanel_10}
\end{figure}

\begin{figure}[H]
    \noindent\makebox[\textwidth]{\includegraphics[width=1.1\textwidth]{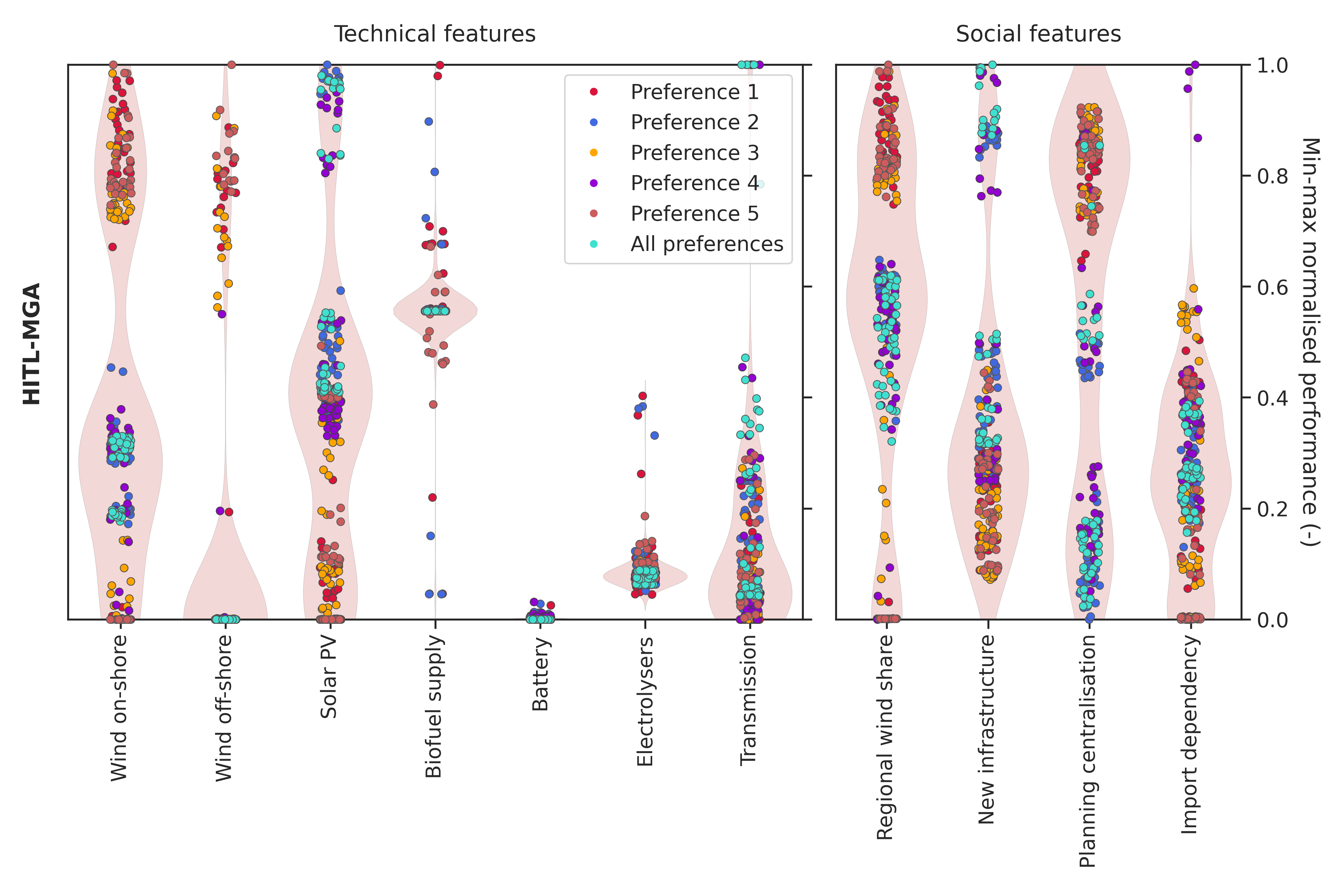}}
    \vspace{-0.5cm}
    \caption{Complements figure \ref{fig:option_spaces_result} by showing the individual contributions of each of the M=(5+1) parallel runs of the HITL-MGA workflow in populating the human-trained option space. As detailed in sub-section \ref{decoding_and_translation}, given M top-rated system designs, the workflow envisions M+1 parallel runs. M runs are each `trained' to target the technical features decoded for one of the top-rated system designs (in this study, M=5); additionally, one more parallel run targets simultaneously all of the non-mutually-exclusive features decoded across all the top-rated system designs.}
    \label{fig:option_spaces_result_per_run}
\end{figure}

\section*{Supporting Methods}

\subsection*{Metrics to assess alignment with high-level preferences}

\begin{table}[H]
	\centering
	\renewcommand{\arraystretch}{1.5} 
        \scriptsize
	\begin{tabularx}{1\textwidth}{
		>{\raggedright\arraybackslash\hsize=.5\hsize}X
		>{\raggedright\arraybackslash\hsize=.5\hsize}X
		>{\raggedright\arraybackslash\hsize=2.5\hsize}X
		>{\raggedright\arraybackslash\hsize=.5\hsize}X
		}
            \hline
            \textbf{Synthetic high-level preferences}                         & \textbf{Assessment metric} & \textbf{Computation logic} & \textbf{Units}     \\
            \hline
            Low import dependency & Import dependency rate & Ratio of electricity and hydrogen imports over total primary energy supply & non-dimensional                                    \\
            Limited wind concentration & Capacity-adjusted wind power share & Linear combination of the highest share of wind power capacity in any model region and the ratio between total deployed wind capacity and the maximum deployed across the original option space & non-dimensional                                   \\
            Limited new infrastructure  & Overall capacity deployment & Total new electricity generation, storage, transmission and electrolysis capacity deployed   & GW                                 \\
            Limited central planning & Degree of deployment of centralised technologies  & One's complement to the weighted sum of capacity deployment for key technologies (the same as those considered for the \textit{Yearly rate of infrastructure deployment}) normalised to their cumulative deployed capacity. The weights indicate qualitatively the extent to which a given technology's deployment can be driven by decentral initiatives. Solar roof-mounted has a decentral weight of 1; battery, wind onshore and solar open-field have a decentral weight of 0.5; all other technologies have a null decentral weight. & non-dimensional                          \\
            Low hydrogen use & Deployment of electrolysers & Deployed electrolysis capacity & GW
            \\
            \hline
            
            \end{tabularx}%
\caption{Summary of metrics used to assess the performance of a given system design option with respect to the assumed synthetic high-level preferences.}
\label{tab:preference_metrics} %
\end{table}

\subsection*{Methods to assign weights in the SPORES MGA algorithm}

In section \ref{decoding_and_translation}, we state that we further differentiate the N=45 system design options generated in each of the M=(5+1) parallel runs of the MGA-HITL workflow by adopting different MGA weights ($w_{ij}$ in Equation \ref{eq:spores_hil}). More precisely, we split these N=45 design options into three parallel batches of N=15 designs each, in which we adopt alternatively \textit{integer}, \textit{relative-deployment} or \textit{evolving-average} weighting methods. These were introduced in prior work \cite{lombardi_what_2023}, and are here reported for convenience.

The \textit{integer} weight-assignment method is reported in Equation \ref{eq:weight-integer}. We apply it to location-technology decision variables ($x_{ij}$) rather than to system-wide technology variables only, thereby making it spatially explicit. The weight is summed to the weight obtained in the preceding iteration ($w_{ij}^{n - 1} $), as it happens for all the other methods below.

\begin{equation} \label{eq:weight-integer}
	w_{ij}^n = w_{ij}^{n - 1} + k_{ij}, \;\; {with}	 \; k_{ij} =
	\begin{cases}	
		100, \; if \; x_{ij}^{cap,n} > c\\
		0, \; if \; x_{ij}^{cap,n} \le c
	\end{cases}
\end{equation}

where $c$ is a threshold defined to avoid very marginal deployments of capacity receiving a weight, which may lead to almost all decision variables receiving a non-zero weight. The weight is set to 100 based on the internal unit scaling of our model and in line with previous work \cite{lombardi_what_2023}.

The \textit{relative-deployment} method, which we devised with a focus on spatial diversity in previous work \cite{lombardi_policy_2020},  is outlined in Equation \ref{eq:weight-relative}.

\begin{equation} \label{eq:weight-relative}
	w_{ij}^n = w_{ij}^{n - 1} + \frac{{x_{ij}^{cap,n}}}{{x_{ij,max}^{cap}}}
\end{equation}

where $x_{ij,max}^{cap}$ is the maximum potential for deployment of a given location-technology decision variable at that location. 

Finally, the \textit{evolving-average} method is illustrated in Equation \ref{eq:weight-evolving}. This method retains a more explicit memory of past iterations. It calculates, for each location-technology decision variable, its (absolute) distance from the average capacity deployed for that variable across all previously found feasible solutions (${\overline{x_{ij}}^{cap,n-1}}$). Such a distance is continuously updated after each iteration. The weight is the inverse of the absolute distance; if the distance is small, the weight is high; vice versa, if the distance is large, the weight is small. The work where we first introduced this method showed that it provides a stronger push for technological diversity compared to the others \cite{lombardi_what_2023}.

\begin{equation} \label{eq:weight-evolving}
	\begin{cases}	
			w_{ij}^n = \left({|\ffrac{ {{\overline{x_{ij}}^{cap,n-1}}} - {x_{ij}^{cap,n}} }{{\overline{x_{ij}}^{cap,n-1}}}|}\right)^{-1}\\
		{{\overline{x_{ij}}^{cap,n-1}}} = \ffrac{\sum\limits_{n=1}^{n-1} x_{ij}^{cap,n}}{n-1}
	\end{cases}
\end{equation}
\microtypesetup{protrusion=false}
\printbibliography
\end{refsection}

\end{document}